\documentclass[prb,onecolumn,showpacs,preprintnumbers,amsmath,assume]{revtex4}
\usepackage{graphicx}	% Include figure files
\usepackage{dcolumn}	% Align table columns on decimal point
\usepackage{bm}			% bold math

\begin{document}

\title{3D two-point ray tracing for heterogeneous, \\ weakly transversely-isotropic media}

\author{Vladimir Grechka$^{1,2}$ and George A. McMechan$^{1}$}
\affiliation{$^{1}$Center for Lithospheric Studies, The University of Texas at Dallas, 
	PO Box 830688, Richardson, TX 75083-0688, USA \\
	$^{2}$currently at Marathon Oil Company}
\date{\today}

\begin{abstract}

A two-point ray-tracing technique for 3D smoothly heterogeneous, weakly transversely-isotropic media is based on Fermat's principle and takes advantage of global Chebyshev approximation of both the model and curved rays. This approximation gives explicit derivatives of traveltimes with respect to ray parameters and allows one to use the rapidly converging conjugate gradient method to compute traveltimes. The method is fast because, for smoothly heterogeneous media, approximation of rays by only a few polynomials and a few conjugate gradient iterations
provide excellent precision of traveltime calculation.

\end{abstract}
   % 81.05.Xj for anisotropy, 91.30.-f for seismology
\maketitle

\section{Introduction} \label{sec:intro}

Elastic anisotropy is widespread in the earth, and many papers discuss the influence of velocity anisotropy on traveltimes and amplitudes of seismic waves. A review of various anisotropic phenomena and an extensive list of references can be found in Crampin and Lovell\cite{CrampinLovell1991}. It is difficult to separate anisotropy from heterogeneity because their influences on seismic wavefields resemble each another. Even shear-wave splitting, commonly attributed to anisotropy, can be caused by strong heterogeneity\cite{GrechkaMcMechan1995}. One way of measuring seismic anisotropy is cross-well traveltime tomography\cite{PrattChapman1992, Michelena1993, Michelenaetal1993} that, under certain conditions, allows one to discriminate the effects of anisotropy and heterogeneity on recorded traveltimes. Because anisotropy is essentially a 3D phenomenon, anisotropic 3D ray tracing is an essential tool for its investigation.

Most of the existing ray-tracing approaches fall into one of three groups: methods based on finite-difference solution of the eikonal equation\cite{Vidale1988, Dellinger1991, vanTrierSymes1991}, shooting methods\cite{CervenyMolotkovPsencik1977, Langanetal1985, GajewskiPsencik1987, VirieuxFarra1991, GuestKendall1993}, and bending methods\cite{UmThurber1987, Protheroetal1988, Schneideretal1992, FariaStoffa1994}, derived, respectively, from Huygens' principle, Snell's law, and Fermat's principle. Methods based on finite-difference solution of the eikonal equation have been applied to 2D anisotropic traveltime calculation\cite{Dellinger1991, QinSchuster1993}. Shooting and bending methods are used for 3D ray tracing\cite{UmThurber1987, Protheroetal1988, VirieuxFarra1991} in isotropic media; only shooting methods have currently been applied for ray tracing in 3D heterogeneous anisotropic media\cite{Cerveny1972, GajewskiPsencik1987, GuestKendall1993}.

We develop a ray-bending technique because two-point ray tracing is more convenient for traveltime tomography than shooting methods. We use global Chebyshev approximation of a heterogeneous anisotropic model and curved rays that make the computations fast. 
 
\section{Methodology} \label{sec:intro}

The main idea of the proposed ray-tracing method is that, in smoothly heterogeneous media, rays are smooth curves that can be approximated by smooth basis functions. We choose Chebyshev polynomials as the basis functions because series of these polynomials usually converge more rapidly to the approximated function than any other polynomial-based series\cite{Lanczos1988}. Fermat's principle is employed for two-point ray tracing. 

To parameterize smoothly heterogeneous media, we use a global 3D approximation of the medium, defined as a sum of Chebyshev polynomials and described by 3D Chebyshev polynomial coefficients or Chebyshev spectral components. The main advantage of this kind of parametrization is that it provides explicit expressions for traveltimes, and, if a ray has already been traced, explicit relations for variations of the traveltime as a function of Chebyshev spectral components of all model and ray parameters. Thus, the traveltime derivatives are obtained at almost no additional computation cost,  allowing us to apply the conjugate gradient method\cite{Pressetal1987} to bend a ray path and minimize the traveltime along it. The absence of the derivatives would entail the use of slower converging minimization methods, such as the Nelder-Mead search\cite{Pressetal1987, Protheroetal1988}.

The proposed technique possesses both the advantages and disadvantages of two-point ray tracing. In some models more than one ray (and traveltime) may exist between a source-receiver pair. Finding all valid solutions is generally difficult because they may correspond not only to minima but also to maxima or saddle points of the traveltime. The same pertains to anisotropy; cusps at the group velocity surfaces, as well as heterogeneity, can produce more than one ray path between the two end points.  Although all these rays may be found by solving special equations\cite{GrechkaObolentseva1993} or by iteration for different initial guesses\cite{ObolentsevaGrechka1988}, we shall restrict our analysis to weakly transversely isotropic media to assure the absence of cusps and the presence of explicit relations for the group velocities\cite{Thomsen1986, Byunetal1989}.

\section{Two-point Chebyshev ray-tracing} \label{sec:chebrt}

\subsection{Model and ray parametrization} \label{sec:chebrt:subsec:mod}

We study kinematics of wave propagation in a heterogeneous transversely isotropic (TI) model occupying  a 3D rectangular volume, specified by the Cartesian coordinates of its corners ${\bm{a}} = (a_1, \, a_2, \, a_3)$ and ${\bm{b}} = (b_1, \, b_2, \, b_3)$. We define the functions
\begin{subequations}
	\begin{align}        
		m_1({\bm{x}}) & \equiv \alpha_0({\bm{x}}), \\
		m_2({\bm{x}}) & \equiv \beta_0({\bm{x}}), \\
 		m_3({\bm{x}}) & \equiv \epsilon({\bm{x}}), \\
		m_4({\bm{x}}) & \equiv \delta  ({\bm{x}}),
 	\end{align}
and
	\begin{align}
		m_5({\bm{x}}) & \equiv \gamma  ({\bm{x}}),
	\end{align}
\end{subequations}
corresponding to Thomsen's\cite{Thomsen1986} anisotropy parameters of TI model, and the directional cosines
\begin{subequations}
	\begin{align}        
	m_6({\bm{x}}) & \equiv c_1({\bm{x}}), \\
	m_7({\bm{x}}) & \equiv c_2({\bm{x}}),
	\end{align}
\end{subequations}
defining the orientation of the unit vector of the symmetry axis 
\begin{subequations}
	\begin{align}        
	{\bm{c}}({\bm{x}}) & \equiv [c_1({\bm{x}}), \, c_2({\bm{x}}), \, c_3({\bm{x}})],
	\end{align}
	where
	\begin{align}
	c_3({\bm{x}}) & = {\sqrt {1 - c_1^2({\bm{x}}) - c_2^2({\bm{x}})}}.
	\end{align}
\end{subequations}
Vector ${\bm{x}}$ in equations (1) and (2) denotes a point ${\bm{x}} \equiv (x_1, \, x_2, \, x_3)$, belonging to the model volume
\begin{equation}
  {\bm{x}} \in [{\bm{a}}, \, {\bm{b}}].
\end{equation}
Parameters $m_{\eta}, ~ (\eta = 1, \ldots, 7)$ specify a weakly TI medium inside the volume. For our objectives, however, it is more convenient to use the 3D Chebyshev spectral components $\mu_{\eta}$ (Appendix A) instead of the functions $m_{\eta} ({\bm{x}})$.

\begin{figure}
	\includegraphics[width = 0.55\textwidth]{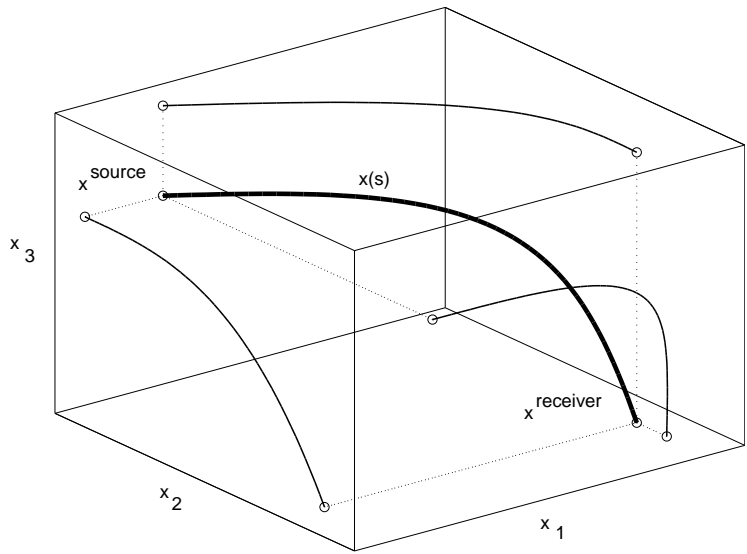} %\vspace{-5mm}
	\caption{A model volume, a ray path (the heavy line) and its projections (the light lines) in 3D Cartesian coordinates.}
	\label{fig01}
\end{figure}

Consider the ray ${\bm{x}}(s)$ connecting a source at ${\bm{x}}^{source}$ and a receiver at ${\bm{x}}^{receiver}$ 
(Figure 1). The Cartesian coordinates $x_i$ of the ray are expanded in the series of Chebyshev polynomials $T_k(s)$
\begin{equation}
  {\bm{x}}(s) \equiv x_i(s) = \sum_{k=1}^{M_i} r_{i,k} \, T_{k-1}(s), \quad (s \in [0,1]; ~ i = 1, \, 2, \, 3),
\end{equation}
where $s$ is the normalized ray length, defined such that
\begin{subequations}
	\begin{align}        
	{\bm{x}}(0) & = {\bm{x}}^{source}
	\end{align}
	and
	\begin{align}
	{\bm{x}}(1) = {\bm{x}}^{receiver},
	\end{align}
\end{subequations}
$r_{i,k}$ are the Chebyshev spectral components of the ray, and $M_i$ are the numbers of the polynomials used to
approximate the ray in each direction $i = 1, \, 2, \, 3$ (Appendix B).

Fermat's principle allows us to compute the Chebyshev spectral components $r_{i,k}$ of a ray.

\subsection{Traveltime computation} \label{sec:chebrt:subsec:ttc}
 
The traveltime along a curved ray connecting the source ${\bm{x}}^{source}$ and the receiver ${\bm{x}}^{receiver}$ is defined as an integral
\begin{equation}
  t_{\rm Q} = \int_{{\bm{x}}^{source}}^{{\bm{x}}^{receiver}} 
p_{\rm Q}(m_{\eta} ({\bm{x}}),\,  {\bm{x}}) \, d {\bm{x}}, \quad ({\rm Q = P, \, SV ~ or ~ SH}),
\end{equation}
where $p_{\rm Q}$ is the group slowness (reciprocal to the group velocity) of P, SV, or SH waves along the ray (Appendix~C), and $m_{\eta}$ are the model parameters [equations (1) and (2)]. Using the ray parametrization given by equations (5) and (6), we rewrite the integral (7) as
\begin{equation}
  t_{\rm Q} = \int_0^1 {\cal T}_{\rm Q}(\mu_{\eta}, \, \bm{r}, \, s) ds,
\end{equation}
where the integrand
\begin{equation}
  {\cal T}_{\rm Q}(\mu_{\eta}, \bm{r}, s) = p_{\rm Q} (m_{\eta}(\bm{r}, \, s), \, \bm{r}, \, s) \, R(\bm{r}, \, s),
\end{equation}
and 
\begin{equation}
  R(\bm{r}, \, s) = \left[ \, \sum_{i=1}^3 \, \dot x_i^2 (s) \right]^{1/2}
\end{equation}
is the length of an element of the ray arc. The dot over a function denotes a derivative with respect to its 
argument, and derivatives $\dot x_i$ are determined by equation (B8) in Appendix B, where the rays are described in terms of Chebyshev polynomials.

Fermat's principle, 
\begin{equation}
  {\partial t_{\rm Q} \over {\partial r_{i,l} } } =
\int_0^1 {\cal D}_{{\rm Q},i,l} (\mu_{\eta}, \, \bm{r}, \, s) \, ds = 0, \quad ({\rm Q = P, \, SV ~ or ~ SH}; ~ i = 1, \, 2, \, 3; ~ l = 3, \, \ldots, \, M_i),
\end{equation}
is used to find the unknown ray spectral coefficients $r_{i,l}$. The index $l$ starts from 3 because the known coordinates of the source and the receiver determine the linear components (for $l = 1, \, 2$) of the ray (see Appendix B for details). The integrand ${\cal D}$ is given by equation (D1) in Appendix D.

Coefficients $r_{i,l}$ that satisfy equations (11) define rays corresponding to minima, maxima, or saddle points of traveltime. For complicated models, the solution of equations (11) is often nonunique and several rays, connecting the same source-receiver pair, may exist. Tracing all such rays is a complicated computational problem because there are no general methods of solving nonlinear systems like equations (11). However, if the solution of system (11) is unique or we want to find only the fastest ray, we can search for the traveltime minimum instead of solving nonlinear equations (11). The conjugate gradient method\cite{Pressetal1987} is applied to find the minimum traveltime. Partial derivatives (11) are the components of the traveltime gradient.

To implement this approach we need to compute integrals (8) and (11). Again, the Chebyshev polynomials are helpful for doing this. Expanding the integrands ${\cal T}$ and ${\cal D}$ as functions of the ray length $s$ in Chebyshev series, we replace the integrals (8) and (11) by the dot products
   \begin{equation}
      t_{\rm Q} = {\bm{P}} \cdot {\bm{C}} ({\cal T}_{\rm Q})
   \end{equation}
and
   \begin{equation}
      {\partial t_{\rm Q} \over {\partial r_{i,l} } } =
                  {\bm{P}} \cdot {\bm{C}} ({\cal D}_{{\rm Q},i,l}),
   \end{equation}
where ${\bm{C}}(\cdot)$ denotes the direct Chebyshev transform of its argument, and vector ${\bm{P}}$ is defined by equation (E6) in Appendix E.

\subsection{Numerical investigation of Chebyshev ray-tracing} \label{sec:chebrt:subsec:nec}
   
We present two numerical examples to illustrate the features of the proposed technique. The efficiency of the method depends on the choice of the two numbers $M$ and $N$. The value of $M$ [equation (B7)] determines the number of unknown Chebyshev spectral coefficients of the ray to be found with the conjugate gradient method. The value of $N$ 
[equation (E1)] defines the number of points needed along each ray to compute traveltime with the required precision. These two quantities are to be determined by numerical experiments.

\begin{figure}
	\includegraphics[width = 0.55\textwidth]{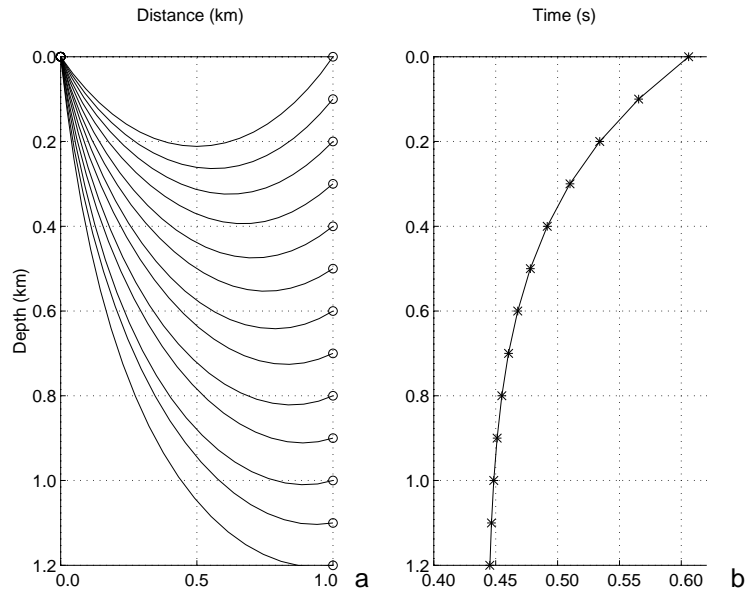} %\vspace{-5mm}
	\caption{The ray trajectories (a) and traveltimes (b) for a source located at $x_i^{source} = (0, \, 0, \, 0)$ and receivers located at $x_i^{receiver} = (x_1, \, 0, \, x_3)$ in a heterogeneous medium described by velocity function $V(x_3) = 1.5 \, \exp (1.5 \, x_3)$ km/s. In (a), circles denote the source and receiver positions; in (b) asterisks denote computed traveltimes, and the solid line denotes the analytic solution computed with equation (15).
	}
	\label{fig02}
\end{figure}

First, we test the method for a model that has a known solution. Consider an isotropic medium with velocity 
\begin{equation}
  V(x_3) = V_0 \, \exp (g \, x_3) ,
\end{equation}
increasing exponentially with depth $x_3$, for parameters $V_0 = 1.5$ km/s and $g = 1.5~{\rm km}^{-1}.$  Velocity function (14) was  approximated by seven Chebyshev polynomials over the depth interval $x_3 \in [0,~ 1.2]$, and traveltimes were computed.  Figure~\ref{fig02} displays the ray trajectories and calculated traveltimes for $M=4$, $N=7$ compared to the analytic solution 
\begin{equation}
  t(x_1, x_3) = { {\sqrt {2 \, ( \cosh g \, x_3 - \cos g \, x_1) } } \over { g \, V_0 \exp (g \, x_3 / 2) } }
\end{equation}
that exists for this velocity model\cite{Goldin1986}. We also performed computations for a set of values $M$ and $N$ listed in Table~\ref{tab1} to study the accuracy of the traveltime calculations and the time required to compute all 13 rays (Figure~\ref{fig02}) on a Sun-4 workstation. Only 1~--~2 iterations of the conjugate gradient method were needed to converge, starting from a straight ray as an initial guess. The number of iterations is found to be independent of $M$ and $N$. Table~\ref{tab1} shows that the computation time depends approximately linearly on $N$ and about quadratically on $M$. Increasing $M$ and $N$ beyond the values presented in Table~\ref{tab1} no longer increases the precision of traveltime calculations because of minor errors associated with the Chebyshev approximation of the velocity function (14). These errors can be made arbitrarily small by using more polynomials to approximate the exponential function in equation (14).

\begin{table}[h]
	\caption{Computation times and accuracy of the traveltime calculations for the
		13 rays in Figure 2 on a Sun-4 workstation, as a function
		of $M$ (the number of Chebyshev polynomials) and $N$ (the number of
		points used in integration).}
	\label{tab1}
	\def\arraystretch{1.11}
	\begin{center}
		\begin{tabular}{||c|c|c|c||} \hline \hline
			 ~   & ~   & Maximum error 		& Computation \\
			 ~ $M$ ~ & ~ $N$ ~ & in traveltime (ms) 	& time (s)    \\ \hline 
			 3 & 4 & 6.46 & 2.13 \\
			   & 5 & 2.80 & 2.79 \\
			   & 6 & 0.45 & 3.06 \\
			   & 7 & 0.12 & 3.76 \\
			   & 8 & 0.12 & 4.05 \\
			   & 9 & 0.12 & 4.90 \\ \hline
			 4 & 4 & 4.82 & 4.04 \\
			   & 5 & 1.41 & 4.50 \\
			   & 6 & 0.37 & 5.34 \\
			   & 7 & 0.10 & 6.01 \\
			   & 8 & 0.05 & 6.75 \\
			   & 9 & 0.02 & 7.65 \\ \hline
			 5 & 4 & 0.98 & 4.85 \\
			   & 5 & 0.81 & 6.75 \\
			   & 6 & 0.06 & 7.50 \\
			   & 7 & 0.04 & 9.16 \\
			   & 8 & 0.02 &10.20 \\
			   & 9 & 0.01 &11.27 \\ \hline \hline
		\end{tabular}
	\end{center}	
	\def\arraystretch{1.0}
\end{table}

The second example is ray tracing in a 3D heterogeneous TI model described by the anisotropy parameters\cite{Thomsen1986} 
\begin{subequations}
	\begin{align}        
	\epsilon & = -0.03 + 0.3 \, x_3, \\
	\delta   & = 0.04 \, (x_1 + x_2),
	\end{align}
	and
	\begin{align}
	\gamma & = 0.05 + 0.02 \, x_1 + 0.03 \, x_2 + 0.04 \, x_3.
	\end{align}
\end{subequations}
The orientation of the symmetry axis is defined by
\begin{subequations}
	\begin{align}        
	c_1 & = 0.5 \, x_3, \\
	c_2 & = 0.5 \, (x_1 - x_2).
	\end{align}
\end{subequations}

\begin{figure}
	\includegraphics[width = 0.75\textwidth]{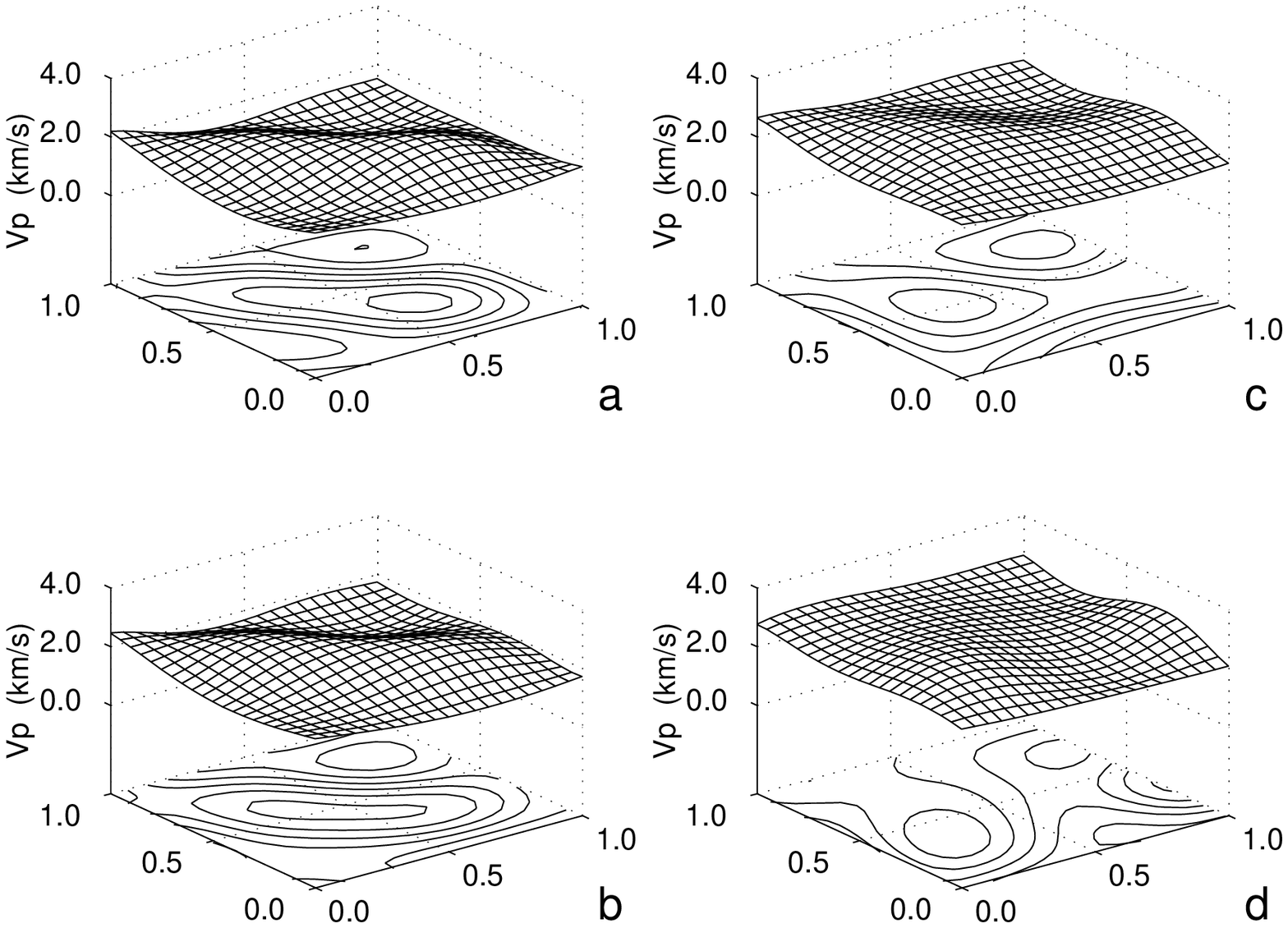} %\vspace{-5mm}
	\caption{Lateral velocity variations $V_P(x_1, \, x_2)$ (in km/s) in the model at depth levels $x_3 = 0.0$ km (a), $x_3 = 0.1$ km (b), $x_3 = 0.2$ km (c), and $x_3 = 0.3$ km (d).
	}
	\label{fig03}
\end{figure}

\begin{figure}
	\centerline{\begin{tabular}{c}
			\includegraphics[width=0.55\textwidth]{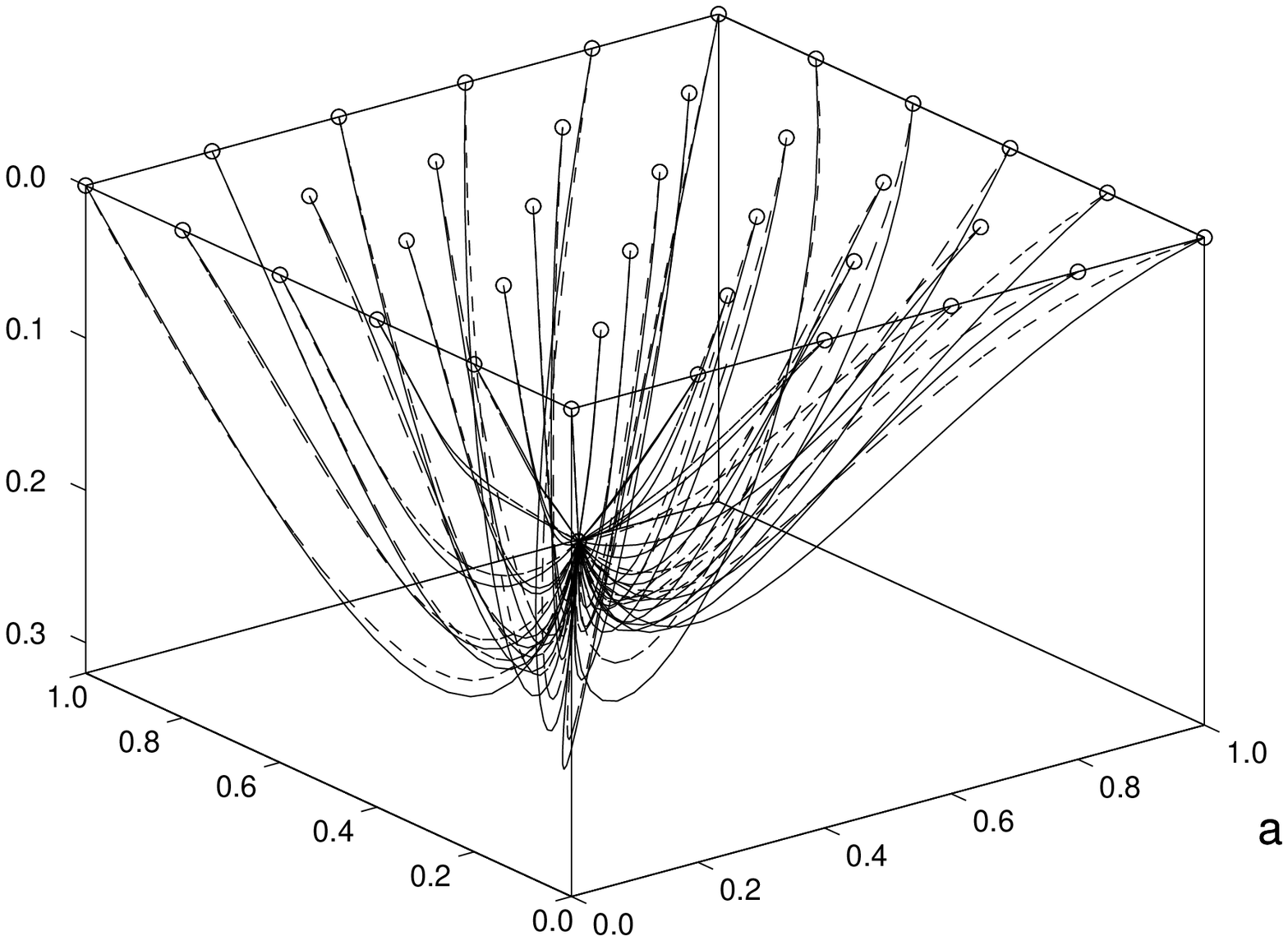} \\
			\includegraphics[width=0.55\textwidth]{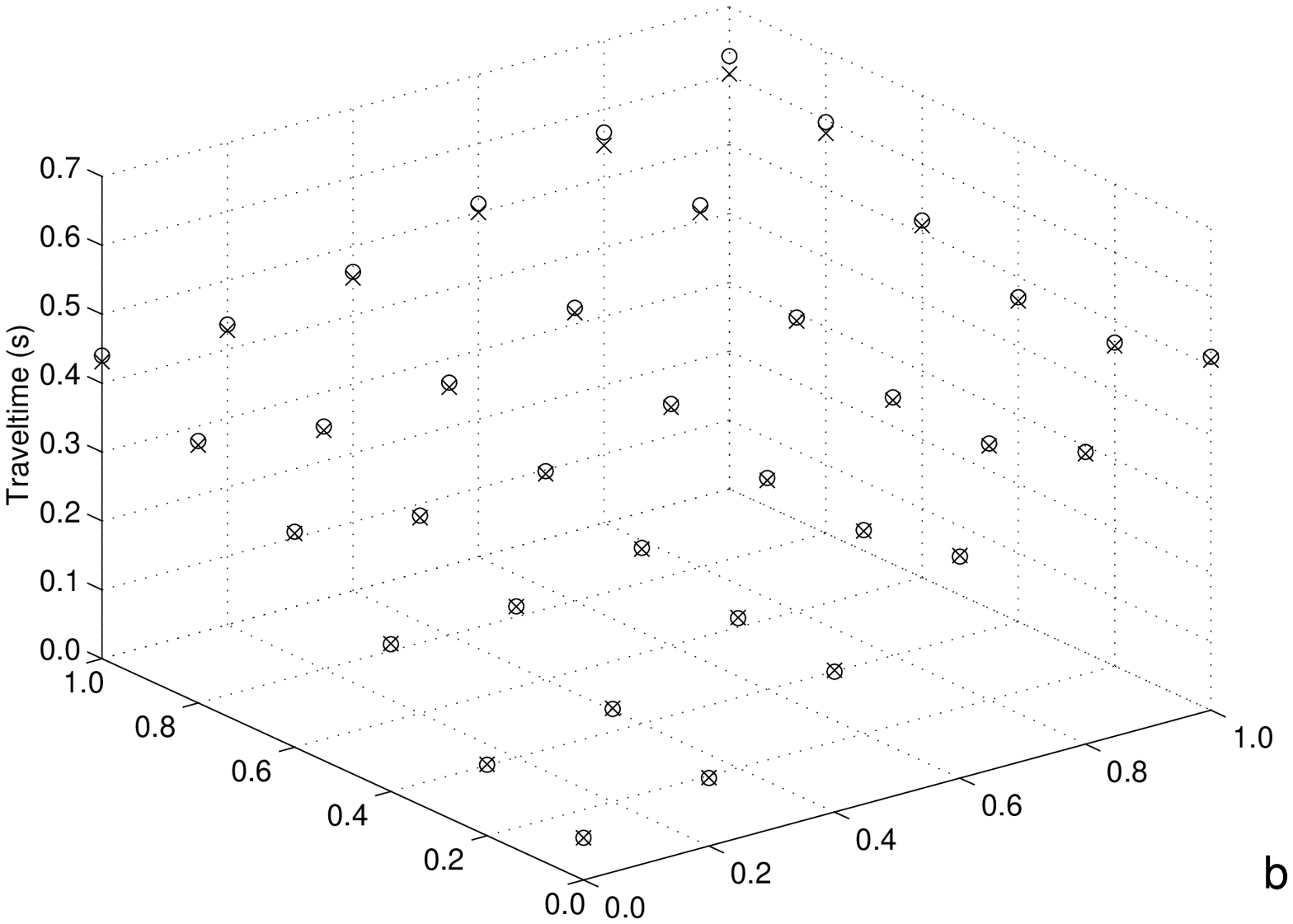} \\
	\end{tabular}} \vspace{0mm}
	\caption{(a) The P-wave ray paths in the anisotropic medium (solid) and in the corresponding isotropic medium (dashed); (b)~computed traveltimes in the anisotropic ($\times$) and isotropic ($\circ$) media.
	}
	\label{fig04} \vspace{0mm}
\end{figure}

\begin{figure}
	\centerline{\begin{tabular}{c}
			\includegraphics[width=0.55\textwidth]{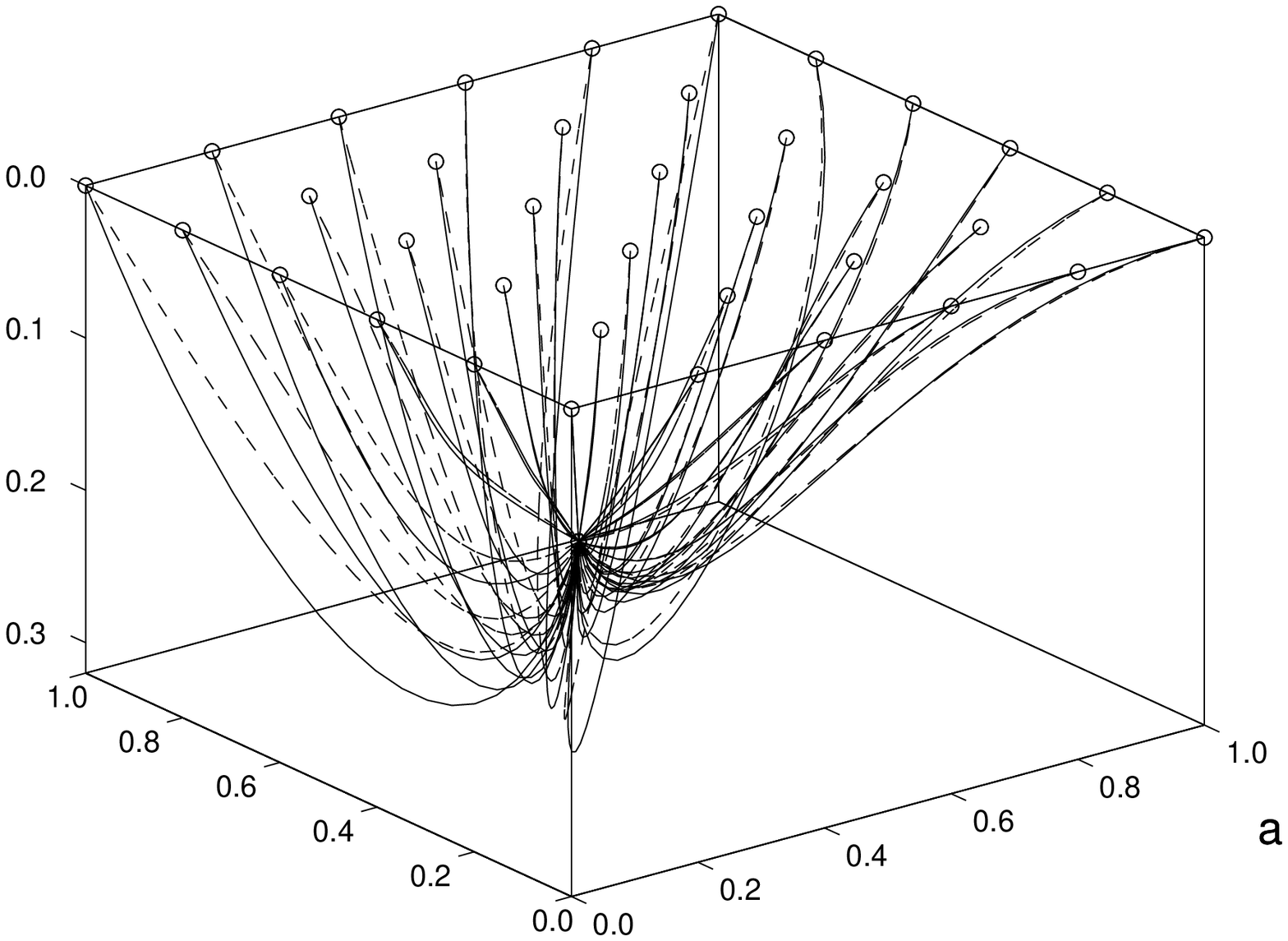} \\
			\includegraphics[width=0.55\textwidth]{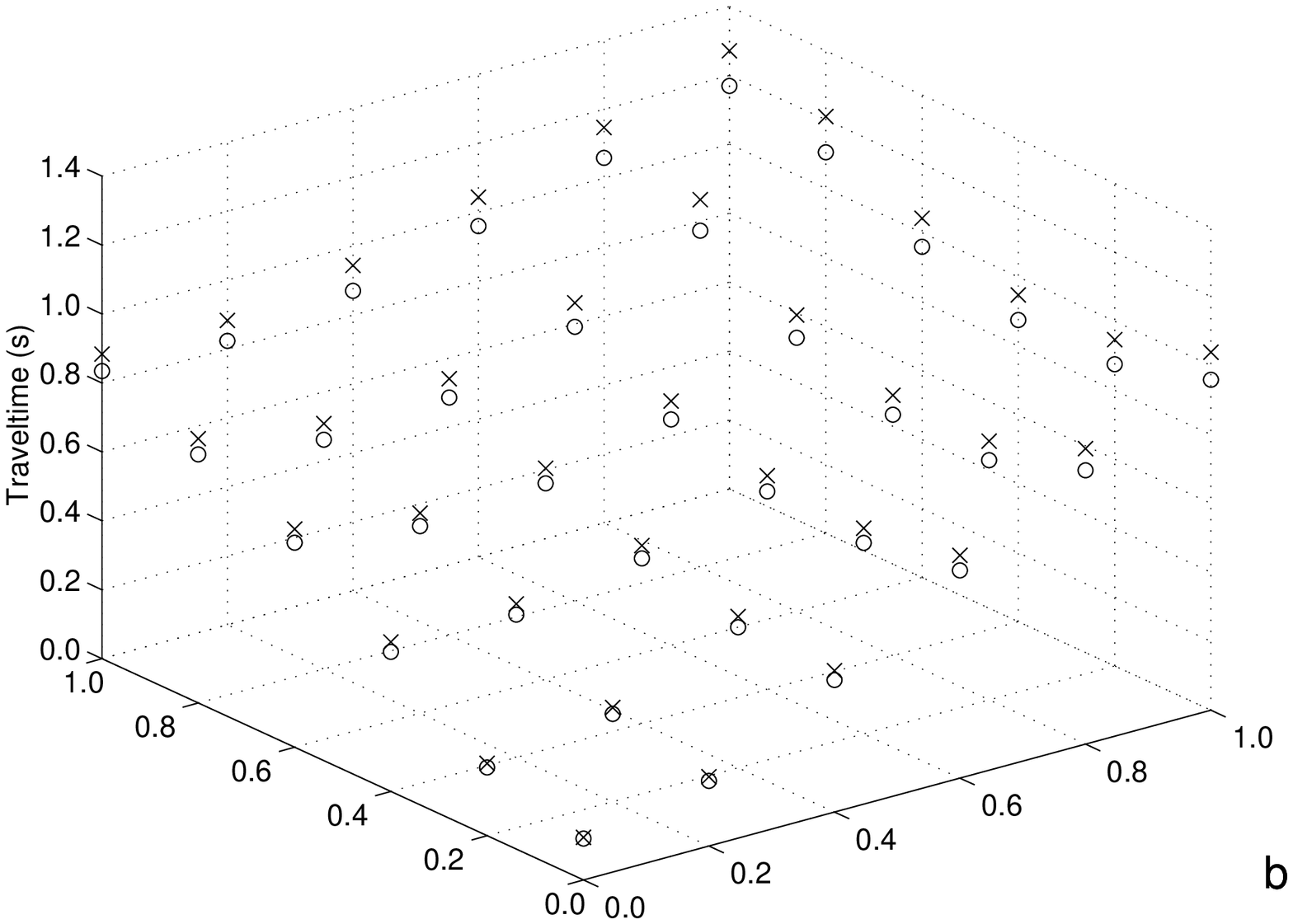} \\
	\end{tabular}} \vspace{0mm}
	\caption{(a) SV (solid) and SH (dashed) ray trajectories; (b) traveltimes of SV ($\times$) and SH ($\circ$) waves. 
	}
	\label{fig05} \vspace{0mm}
\end{figure}

The 3D distribution of the P-wave velocity $\alpha_0({\bm{x}})$ is shown in Figure~\ref{fig03}, whereas the shear velocity is defined as \mbox{$\beta_0({\bm{x}}) = \alpha_0({\bm{x}})/2$} [equations (C1) and (C4)]. The other model functions, influencing the traveltimes to a lesser degree, are not shown. The source coordinates (in km) are  ${\bm{x}}^{source} = [0.05, \, 0.05, \, 0.1]$, and 36 receivers are located at the coordinates ${\bm{x}}^{receiver} = [0.2 \, i, \, 0.2 \, j, \, 0.0], ~ (i, \, j = 0, \, \ldots, \, 5)$; these are displayed in Figures~\ref{fig04}a and~\ref{fig05}a.

We use $M_1 = M_2 = M_3 = 5$ polynomials to approximate the rays (i.\,e., 9 unknown spectral components for each ray are to be found) and $N=9$ points along the rays to perform integration. The conjugate gradient method converges after 2~--~3 iterations. Comparing the results with those in the first example, we expect the accuracy of our traveltime calculations to be around $10^{-4}$ s or better.

Figure~\ref{fig04}a presents the P-wave ray trajectories in the specified TI medium and the trajectories in the related isotropic medium, for which $\epsilon = \delta = 0$ in equations (16a) and (16b). Corresponding traveltimes are shown in Figure~\ref{fig04}b. Although the ray paths differ substantially in the anisotropic and isotropic models, the traveltimes do not.

A similar comparison of ray trajectories and traveltimes for SV and SH waves in the same TI model is given in Figure~\ref{fig05}. For this model, the heterogeneity influences traveltimes more than the anisotropy does.

\section{Discussion} \label{sec:disc}

Two-point ray tracing in 3D  heterogeneous anisotropic media may be based on Chebyshev approximation of curved rays.
The method is inexpensive in smoothly varying media because only a few polynomials are needed to approximate the ray paths. The global Chebyshev model parametrization provides explicit relations for partial derivatives of traveltimes as functions of the ray parameters. This allows us to use the rapidly converging conjugate gradient method to compute the traveltimes.

The computing time depends directly on the values of $M$ (the number of Chebyshev polynomials approximating a ray) and $N$ (the number of integration points per ray). These quantities were determined in the first example by comparing the traveltimes to the known solution. Generally, when the solution is not known, $M$ and $N$ should be determined by numerical experiments. A few representative rays and traveltimes $t(M,N)$ are computed for a set of values $M$ and $N$. As $M$ and $N$ become greater traveltimes usually gradually decrease, whereas the computation time increases, as illustrated in Table~\ref{tab1}. We choose $M$ and $N$ that correspond to the minimum computation time for which traveltimes are about 0.05\% greater than the minimum $t(M,N)$ for all tested pairs of $M$ and $N$. The values of $M=5$ and $N=9$ were selected by this procedure for the second example.

Although we applied only the conjugate gradient method to compute the ray trajectories and traveltimes, other methods of solving this problem are available. For example, one could compute the Hessian matrix $\partial^2 t / \partial \bm{r}^2$ and use the second order Newton method, which generally reduces number of iterations.

The proposed technique is fast for weakly TI media, where the group velocity explicitly depends on a ray direction. If anisotropy is not weak, we would have to compute the group velocity for a given ray direction numerically. This can be done iteratively\cite{ObolentsevaGrechka1988} and involves searching for one phase angle for TI media and for two phase angles for more generally anisotropic media. This search, done for every evaluation of the group velocity, would significantly increase the computational time. On the other hand, the proposed technique applied to 3D isotropic media is faster by a factor of about five, as becomes clear from the examination of equations (C4) for the group velocities. To find the velocities, some of the quantities $m_j ~ (j = 1, \, \ldots, \, 7)$ have to be computed with equation (A4), which takes most of the computational effort. For an isotropic medium though, only the medium velocity itself needs to be computed.

It is expected that increasing the complexity of the model (especially when functions describing the model parameters are discontinuous) would reduce the computational efficiency. Although Chebyshev ray tracing can be applied to models with discontinuities, the number of polynomials (and perhaps the number of required iterations) would no longer be small, and the technique loses its elegance. Another approach for discontinuous parameters is to separate the medium into blocks along surfaces of discontinuity (interfaces) and to construct rays as sequences of their smooth segments. This idea seems to be feasible but lies beyond the scope of this paper.

\section{Acknowledgments} \label{sec:acknowl}

The research leading to this paper was supported by a UT-Dallas graduate teaching assistantship (V.G.) and by the sponsors of the UTD Geophysical Consortium (G.A.M.). The computations were performed on a Sun-4 workstation in the Center for Lithospheric Studies at the University of Texas at Dallas. V.G. wishes to thank Sergey Fomel for the idea to perform
the Chebyshev ray tracing in 3D rather than in 2D and Ilya Tsvankin for helpful discussions. We thank to Spyros Lazaratos, Reinaldo Michelena and  an anonymous reviewer whose comments and suggestions improved the manuscript. This paper is Contribution No. 825 from the Department of Geosciences at the University of Texas at Dallas.

\appendix
 
\section{3D Chebyshev transform}
\label{app:first}

The 3D Chebyshev transform is constructed as a generalization of the 1D Chebyshev transform\cite{Lanczos1988}. A vector   
${\bm{x}} \equiv (x_1, \, x_2, \, x_3)$ belonging to the volume ${\bm{x}} \in [{\bm{a}}, {\bm{b}}]$ [equation (4)] can be transformed to the unit cube ${\bm{y}} \in [\bm{0}, \, \bm{1}]$ by the linear transformation
   \begin{equation}
      y_i = 1 + { { x_i - b_i} \over {b_i - a_i} }, \quad (i = 1, \, 2, \, 3).
   \end{equation}
The corresponding inverse transform is
   \begin{equation}
      x_i = a_i + (b_i - a_i) y_i, \quad (i = 1, \, 2, \, 3).
   \end{equation}

Let us define a set of orthonormal Chebyshev polynomials
   \begin{equation}
      T_{{\bm{k}}-1} (\bm{y}) \equiv T_{k_i - 1} (y_i), \quad (i = 1, \, 2, \, 3; ~ k_i = 1, \, \ldots, \, N_i)
   \end{equation}
inside the cube ${\bm{y}} \in [\bm{0}, \, \bm{1}]$ and express a function $m({\bm{x}})$ as the 3D inverse Chebyshev transform
   \begin{equation}
      m({\bm{x}}) \equiv m(x_1, \, x_2, \, x_3) =
	 \sum_{k_1=1}^{N_1} \sum_{k_2=1}^{N_2} \sum_{k_3=1}^{N_3} 
	 \biggl( \mu (k_1, \, k_2, \, k_3) \prod_{i=1}^3 [T_{k_i - 1} (y_i)] \biggr).
   \end{equation}

The 3D Chebyshev spectral components $\mu({\bm{k}})$ are found by the 3D
direct Chebyshev transform
   \begin{equation}
      \mu ({\bm{k}}) \equiv \mu(k_1, \, k_2, \, k_3) =
	 {\biggl( \prod_{i=1}^3 N_i \biggr) }^{\!\!-1}
	 \sum_{j_1=1}^{N_1} \sum_{j_2=1}^{N_2} \sum_{j_3=1}^{N_3} 
	 \biggl( m(\chi_{j_1}, \, \chi_{j_2}, \, \chi_{j_3}) 
	    \prod_{i=1}^3 [T_{k_i - 1} (\lambda_{j_i})] \biggr),
   \end{equation}
where $\lambda_{j_i}$ are the roots of the Chebyshev polynomials
   \begin{equation}
      \lambda_{j_i} = {1 \over 2} \, \left( 1 + \cos {{2 \, j_i - 1} \over {2 \, N_i}} \pi \right),
   \end{equation}
and $\chi_{j_i}$ relate to $\lambda_{j_i}$ through equation (A2), as
   \begin{equation}
      \chi_{j_i} = a_i + (b_i - a_i) \lambda_{j_i}, \quad (i = 1, \, 2, \, 3; ~ j_i = 1, \, \ldots, \, N_i).
   \end{equation}

\section{Rays and its derivatives} \label{app:second}

The Chebyshev polynomial approximation of a ray ${\bm{x}}(s)$ is defined by equation (5) as
   \begin{equation}
      x_i(s) = \sum_{k=1}^{M_i} r_{i,k} \, T_{k-1}(s), \quad (s \in [0, \, 1]; ~ i = 1, \, 2, \, 3).
   \end{equation}
The source and receiver conditions (6) give six constraints on components $r_{i,k}$
\begin{subequations}
	\begin{align}        
	 x_i^{source} = x_i(0) = \sum_{k=1}^{M_i} r_{i,k} \, T_{k-1}(0), 
	\end{align}
	and
	\begin{align}
	 x_i^{receiver} = x_i(1) = \sum_{k=1}^{M_i} r_{i,k} \, T_{k-1}(1), \quad (i = 1, \, 2, \, 3) ,
	\end{align}
\end{subequations}
rewritten as
\begin{subequations}
	\begin{align}        
	x_i^{source} = r_{i,1} - {\sqrt 2} \, r_{i,2} + {\sqrt 2} \, \sum_{k=3}^{M_i} (-1)^{k-1} r_{i,k}, 
	\end{align}
	and
	\begin{align}
	x_i^{receiver} = r_{i,1} + {\sqrt 2} \, r_{i,2} + \sum_{k=3}^{M_i} \, r_{i,k}, \quad (i = 1, \, 2, \, 3),
	\end{align}
\end{subequations}
using values of the Chebyshev polynomials at the edges of the segment $[0, \, 1]$
\begin{subequations}
	\begin{align}        
	{\bm{T}} (0) = [1, \, -{\sqrt 2}, \, {\sqrt 2}, \, -{\sqrt 2}, \, \ldots]
	\end{align}
	and
	\begin{align}
	{\bm{T}} (1) = [1, \, {\sqrt 2}, \, {\sqrt 2},  \, {\sqrt 2}, \, \ldots]
	\end{align}
\end{subequations}
as the boundary conditions. The components $r_{i,1}$ and $r_{i,2}$ are found from equations (B3) and substituted into equation (B1) to give
   \begin{equation}
      x_i(s) = (1-s) \, x_i^{source} + s x_i^{receiver} +
               \sum_{k=3}^{M_i} \biggl[ {\sqrt 2} \, \big( (1-s) \, (-1)^k - s \big) + T_{k-1}(s) \biggr] r_{i,k},
   \end{equation}
where the relations
\begin{subequations}
	\begin{align}        
	T_0 (s) = 1
	\end{align}
	and
	\begin{align}
	 T_1 (s) = {\sqrt 2} \, (-1 + 2 \, s) 
	\end{align}
\end{subequations}
were used. Equation (B5) shows that the ray ${\bm{x}}(s)$ is specified by 
   \begin{equation}
      M = M_1 + M_2 + M_3 - 6
   \end{equation}
quantities $r_{i,k}$, given that the source and receiver conditions (B2) are satisfied.

All the required derivatives can be found by explicit differentiation of equation (B5). These derivatives are
   \begin{equation}
      {\dot x}_i(s) = x_i^{receiver} - x_i^{source} + \sum_{k=3}^{M_i} \left[ {\dot T}_{k-1}(s) - 
					 {\sqrt 2} \, (1 - (-1)^{k-1}) \right] \, r_{i,k},
   \end{equation}
   \begin{equation}
      { {\partial x_i(s)} \over {\partial r_{i,l}} } = -{\sqrt 2} \left[ (1-s) \, (-1)^{l-1} + s \right] + T_{l-1}(s),
   \end{equation}
and
   \begin{equation}
      { {\partial {\dot x}_i(s)} \over {\partial r_{i,l}} } = -{\sqrt 2} \left[ 1 - (-1)^{l-1} \right] + {\dot T}_{l-1}(s), \quad (i = 1, \, 2, \, 3; ~ l = 3, \, \ldots, \, M_i).
   \end{equation}
The dot over the functions in equations (B8) and (B10) denotes a derivative with respect to the argument.

\section{The group slowness in weakly TI and its derivatives} \label{app:third}

Phase velocities of the P, SV, and SH waves as functions of angle $\theta$ between the wavefront normal and the symmetry axis are given by\cite{Thomsen1986} 
\begin{subequations}
	\begin{align}        
	v_{\rm P} (\theta) & = \alpha_0 \, (1 + \delta \, \sin^2 \theta \, \cos^2 \theta + \epsilon \, \sin^4 \theta ), \\
	v_{\rm SV} (\theta) & = \beta_0 \biggl[1 + {\alpha_0^2 \over \beta_0^2} \, (\epsilon - \delta) \, \sin^2 \theta \, \cos^2 \theta \biggr],
	\end{align}
	and
	\begin{align}
	v_{\rm SH} (\theta) & = \beta_0 (1 + \gamma \, \sin^2 \theta ).
	\end{align}
\end{subequations}
Equations (C1), valid for small $\epsilon$, $\delta$, and $\gamma$ (i.e., for weak anisotropy), provide the linear approximation of the phase velocities as functions of $\epsilon$, $\delta$, and $\gamma$. Also the group velocity equals the phase velocity in the linear approximation\cite{Thomsen1986}. Thus,
\begin{subequations}
	\begin{align}        
	V_{\rm P} (\phi) & = v_{\rm P} (\theta), \\
	 V_{\rm SV} (\phi) & = v_{\rm SV} (\theta),
	\end{align}
	and
	\begin{align}
	V_{\rm SH} (\phi) & = v_{\rm SH} (\theta),
	\end{align}
\end{subequations}
where $\phi$ is an angle between a ray and the symmetry axis.

In the linear approximation, the relation between the angles $\theta$ and $\phi$ reads\cite{Thomsen1986}
   \begin{equation}
      \theta = \phi + \Delta,
   \end{equation}
where $\Delta$, being different for P, SV, and SH waves, is a small quantity that has the same order as $\epsilon$, $\delta$, and $\gamma$. Substituting equations (C1) and (C3) into equation (C2) and neglecting quadratic terms in $\epsilon$, $\delta$, $\gamma$, and $\Delta$, we obtain the group velocities in the linear approximation as
\begin{subequations}
	\begin{align}        
	V_{\rm P} (\phi) & = \alpha_0 \, (1 + \delta \, \sin^2 \phi \, \cos^2 \phi + \epsilon \, \sin^4 \phi ), \\
	V_{\rm SV} (\phi) & = \beta_0 \biggl[1 + {\alpha_0^2 \over \beta_0^2} \, (\epsilon - \delta) \, \sin^2 \phi \, \cos^2 \phi \biggr],
	\end{align}
	and
	\begin{align}
	V_{\rm SH} (\phi) = \beta_0 (1 + \gamma \, \sin^2 \phi ).
	\end{align}
\end{subequations}
The same approximation for the group slownesses gives
\begin{subequations}
	\begin{align}        
	p_{\rm P} (\psi) & = {1 \over V_{\rm P} (\phi)} = {1 \over \alpha_0} \biggl[1 - \epsilon + (2 \epsilon - \delta) \, \psi + (\delta - \epsilon ) \, \psi^2 \biggr], \\
	p_{\rm SV} (\psi) & = {1 \over V_{\rm SV} (\phi)} = {1 \over \beta_0} \biggl[1 - {\alpha_0^2 \over \beta_0^2} \, 	(\epsilon - \delta) \, \psi \, (1 - \psi) \biggr],
	\end{align}
	and
	\begin{align}
	p_{\rm SH} (\phi) = {1 \over V_{\rm SH} (\phi)} = {1 \over \beta_0} \biggl[ 1 - \gamma \, (1 - \psi) \biggr],
	\end{align}
\end{subequations}
where
   \begin{equation}
      \psi = \cos^2 \phi.
   \end{equation}
The approximation for the square of the group slowness, in a form similar to equation (C5a), was derived earlier\cite{Byunetal1989}.

The quantity $\psi$ is the square of the dot product of vector ${\bm{c}}$ [equations (2) and (3)] and the normalized tangent to the ray $\dot {\bm{x}}/R$ [equations (B8) and (9)];
\begin{equation}
   \psi = \biggl( {\bm{c}} \cdot {\dot {\bm{x}} \over R} \biggr)^{\!2} = {\Phi^2 \over R^2} \, ,
\end{equation}
where 
\begin{equation}
   \Phi = ({\bm{c}} \cdot \dot {\bm{x}} ) = c_1 \, {\dot x_1} + c_2 \, {\dot x_2} + {\sqrt {1 - c_1^2 - c_2^2} } \, \, {\dot x_3} \, .
\end{equation}

Our next task is to calculate the derivatives $\partial p_{\rm Q} / \partial m_\eta$, where Q = P, SV, or SH, and $m_\eta$ are defined by equations (1), (2). Differentiating equations (C5), we obtain
\begin{subequations}
	\begin{align}        
	{\partial p_{\rm P} \over \partial m_1} & \equiv {\partial p_{\rm P} \over \partial \alpha_0} = 
	- {p_{\rm P} \over \alpha_0}, \\
	 {\partial p_{\rm P} \over \partial m_2} & \equiv {\partial p_{\rm P} \over \partial \beta_0} = 0, \\
	{\partial p_{\rm P} \over \partial m_3} & \equiv {\partial p_{\rm P} \over \partial \epsilon} = 
	- { (1-\psi)^2 \over \alpha_0}, \\
	{\partial p_{\rm P} \over \partial m_4} & \equiv {\partial p_{\rm P} \over \partial \delta} = 
	- { \psi (1-\psi) \over \alpha_0}, \\
	{\partial p_{\rm P} \over \partial m_5} & \equiv {\partial p_{\rm P} \over \partial \gamma} = 0, \\
	{\partial p_{\rm P} \over \partial m_{5+l}} & \equiv {\partial p_{\rm P} \over \partial c_l} = 
	{\partial p_{\rm P} \over \partial \psi} \, {\partial \psi \over \partial c_l} \, , \quad (l = 1, \, 2), \\
	{\partial p_{\rm SV} \over \partial m_1} & \equiv {\partial p_{\rm SV} \over \partial \alpha_0} = 
	- 2 \, {\alpha_0  \over \beta_0^3} \, (\epsilon - \delta) \, \psi \, (1 - \psi), \\
	{\partial p_{\rm SV} \over \partial m_2} & \equiv {\partial p_{\rm SV} \over \partial \beta_0} = 
	- {1 \over \beta_0^2} + 4 \, {\alpha_0^2  \over \beta_0^4} \, (\epsilon - \delta) \, \psi \, (1 - \psi), 
	\end{align}
	\begin{align}
	{\partial p_{\rm SV} \over \partial m_3} & \equiv {\partial p_{\rm SV} \over \partial \epsilon} = 
	- {\alpha_0^2  \over \beta_0^3} \, \psi \, (1 - \psi), \\
	{\partial p_{\rm SV} \over \partial m_4} & \equiv {\partial p_{\rm SV} \over \partial \delta} = 
	{\alpha_0^2  \over \beta_0^3} \, \psi \, (1 - \psi), \\
	{\partial p_{\rm SV} \over \partial m_5} & \equiv {\partial p_{\rm SV} \over \partial \gamma} = 0, \\
	{\partial p_{\rm SV} \over \partial m_{5+l}} & \equiv {\partial p_{\rm SV} \over \partial c_l} = 
	{\partial p_{\rm SV} \over \partial \psi} \, {\partial \psi \over \partial c_l} \, , \quad (l = 1, \, 2), \\
	{\partial p_{\rm SH} \over \partial m_1} & \equiv {\partial p_{\rm SH} \over \partial \alpha_0} = 0, \\
	{\partial p_{\rm SH} \over \partial m_2} & \equiv {\partial p_{\rm SH} \over \partial \beta_0} = 
	- {p_{\rm SH} \over \beta_0}, \\
	{\partial p_{\rm SH} \over \partial m_3} & \equiv {\partial p_{\rm SH} \over \partial \epsilon} = 0, \\
	{\partial p_{\rm SH} \over \partial m_4} & \equiv {\partial p_{\rm SH} \over \partial \delta} = 0, \\
	{\partial p_{\rm SH} \over \partial m_5} & \equiv {\partial p_{\rm SH} \over \partial \gamma } = 
	{\psi -1 \over \beta_0},
	\end{align}
	and
	\begin{align}
	{\partial p_{\rm SH} \over \partial m_{5+l}} & \equiv {\partial p_{\rm SH} \over \partial c_l} = 
	{\partial p_{\rm SH} \over \partial \psi} \, {\partial \psi \over \partial c_l} \, , \quad (l = 1, \, 2),
	\end{align}
\end{subequations}
where
\begin{subequations}
	\begin{align}        
	{\partial p_{\rm P} \over \partial \psi} & = {1 \over \alpha_0} \biggl[ (2 \, \epsilon - \delta) + 2 \, \psi \, (\delta - \epsilon ) \biggr], \\
	{\partial p_{\rm SV} \over \partial \psi} & = -{\alpha_0^2 \over \beta_0^2} \, (\epsilon - \delta) \, (1 - 2 \, \psi), \\
	{\partial p_{\rm SH} \over \partial \psi} & = {\gamma \over \beta_0} \, ,
	\end{align}
	and
\end{subequations}
\begin{equation}
	{\partial \psi \over \partial c_l} = {2 \, \Phi \over R^2} \biggl[ \dot x_l - { {c_l \, \dot x_3} \over {\sqrt {1 - c_1^2 - c_2^2} } } \biggr]. \quad (l = 1, \, 2),
\end{equation}

Because the group slowness (C5) depends directly on $\psi$, which in turn, depends on the Chebyshev spectral components of the ray [equations (10), (B1), (C7) and (C8)], the slowness also depends on these components. The corresponding partial derivatives are
   \begin{equation}
      {\partial p_{\rm Q} \over \partial r_{i,l} } = {\partial p_{\rm Q} \over \partial \psi} \,
         {\partial \psi \over \partial r_{i,l} },
   \end{equation}
where 
   \begin{equation}
      {\partial \psi \over \partial r_{i,l} } = { {2 \, \Phi} \over R^3} 
	 \biggl[ R \, {\partial \Phi \over \partial r_{i,l} } - \Phi \, {\partial R \over \partial r_{i,l} } \biggr],
   \end{equation}
   \begin{equation}
      {\partial R \over \partial r_{i,l} } = { {\dot x_i} \over R} \, {\partial {\dot x_i} \over \partial r_{i,l} },
   \end{equation}
   \begin{equation}
      {\partial \Phi \over \partial r_{i,l} } = c_i \, {\partial {\dot x_i} \over \partial r_{i,l} } \, , \quad (i = 1, \, 2, \, 3; ~ l = 1, \, \ldots, \, M_i);
   \end{equation}
derivatives ${\partial p_{\rm Q} / \partial \psi}$ are given by equations (C10), derivatives ${\dot x_i}$ and 
${\partial {\dot x_i} / \partial r_{i,l} }$ -- by equations (B8) and (B10).

\section{Computation of integrand ${\cal D}$ in equation (11)} \label{app:forth}

To compute the integrand ${\cal D}_{{\rm Q},i,l} = \partial {\cal T_{\rm Q}} / \partial r_{i,l}$ in equation (11) for anisotropic heterogeneous media, consider the following. First, the group slowness $p_{\rm Q}$ in equation (9) depends on the model parameters $m_{\eta}$ that depend on the spectral components $r_{i,l}$, because the medium is heterogeneous. Second, the group slowness also depends directly on $r_{i,l}$ because of anisotropy (i.e., different $r_{i,l}$ define different ray directions and accordingly different values of the slowness at the same point). Third, the components $r_{i,l}$ are constrained by six equations (B2), determining the components $r_{i,1}$ and $r_{i,2}$. Therefore, we have to find the derivatives $\partial {\cal T_{\rm Q}} / \partial r_{i,l}$ for $l > 2.$

Differentiating equation (9) yields
   \begin{equation}
      {\cal D}_{{\rm Q},i,l} \equiv { {\partial {\cal T_{\rm Q}}} \over {\partial r_{i,l}} } =
	 R \, \Biggl[ \sum_{\eta=1}^7 \biggl( { {\partial p_{\rm Q}} \over {\partial m_{\eta}} } \,
		   { {\partial m_{\eta}} \over {\partial r_{i,l}} } \biggl)
           + { {\partial p_{\rm Q}} \over {\partial r_{i,l}} } \Biggr]
        + p_{\rm Q} \, { {\partial R } \over {\partial r_{i,l}} }, 
   \end{equation}
\centerline{$({\rm Q = P, SV ~ or ~ SH}; ~ i = 1, \, 2, \, 3; ~ l = 1, \, \ldots, \, M_i),$} 

\noindent where the summation from 1 to 7 and the presence of the second term in the brackets are due to anisotropy. Derivatives ${ {\partial p_{\rm Q}} / {\partial m_{\eta}} }$ in (D1) are given by equations (C9), derivatives 
${ {\partial p_{\rm Q}} / {\partial r_{i,l}} }$ by equation (C12), and derivatives ${ {\partial R } / {\partial r_{i,l}} }$ by equation (C14).

The derivatives of the model parameters $m_{\eta}({\bm{x}})$, obtained from equations (A4), are
   \begin{equation}
      { {\partial m_{\eta}({\bm{x}})} \over {\partial r_{i,l}} } =
      { 1 \over {b_i - a_i} } \sum_{k_1=1}^{N_1^{(\eta)}} \sum_{k_2=1}^{N_2^{(\eta)}} \sum_{k_3=1}^{N_3^{(\eta)}}
	    \biggl( \mu_{\eta} (k_1, \, k_2, \, k_3) \biggl[ \prod_{j=1 \atop j \ne i}^3 T_{k_j -1} (y_j) \biggr] 
	       \dot T_{k_i-1} (y_i) \, { {\partial x_i} \over {\partial r_{i,l}} } \biggr),
   \end{equation}
\centerline{$ (\eta = 1, \, \ldots, \, 7; ~ i = 1, \, 2, \, 3; ~ l = 1, \ldots, M_i), $}

\noindent where $N_i^{(\eta)}$ are the numbers of polynomials approximating the $\eta^{\rm th}$ parameter along $i^{\rm th}$ coordinate axis, and derivatives ${ {\partial x_i} / {\partial r_{i,l}} }$ are defined by equation (B9).

\section{The Chebyshev integration} \label{app:fifth}

Consider the 1D direct Chebyshev transform of a function $f(z)$ at the interval $z \in [0, \, 1]$ 
   \begin{equation}
      {\bm{C}} (f) \equiv f_k = {1 \over N} \sum_{j=1}^N f(\lambda_j) \, T_{k-1}(\lambda_j),
   \end{equation}
where ${\bm{C}} (f)$ is the symbol of the direct Chebyshev transform of function $f$ and the roots of the Chebyshev polynomials $\lambda_j$ are given by equation (A6).

Integration in the Chebyshev domain reduces to a simple multiplication by the integral matrix ${\bm{J}}$,
   \begin{equation}
      {\bm{C}} \biggl( \int_0^z f(\xi) \, d \xi \biggr) = {\bm{J}} \, {\bm{C}} (f) \equiv \sum_{k=1}^N J_{ik} \, f_k, 
      \quad (0 \le z \le 1; ~ i = 1, \, \ldots, \, N),
   \end{equation}
where the matrix ${\bm{J}}$ is\cite{Volovodenko1981} 
   \begin{equation}
	 {\bm{J}} = {1 \over 4} 
	 \begin{pmatrix}
	 2 & -\dfrac{\sqrt{2}}{2} & -\dfrac{2\sqrt{2}}{3} & \ldots &
             \dfrac{2\sqrt{2} (-1)^{N-1}}{(N-2)^2-1} & \dfrac{2\sqrt{2} (-1)^{N}}{(N-1)^2-1} \cr
	  \sqrt{2} &      0      &    -1    & \ldots &    0    &    0 \cr
	      0    & \dfrac{1}{2} &     0    & \ldots &    0    &    0 \cr
            \vdots &   \vdots    &  \vdots  & \ddots &  \vdots & \vdots \cr
              0    &      0      &     0    & \ldots &    0 
						     & -\dfrac{1}{N-2} \cr
	      0    &      0      &     0    & \ldots & \dfrac{1}{N-1} & 0 \cr
   \end{pmatrix} .
   \end{equation}

To evaluate the integral in equation (E2), we apply the inverse Chebyshev transform of the vector ${\bm{J}} \, {\bm{C}}(f).$ If integral at only one point $z$ is needed, the inverse transform becomes a single dot product
   \begin{equation}
      \int_0^z f(\xi) \, d \xi = {\bm{T}} (z) \cdot {\bm{J}} \, {\bm{C}} (f),
   \end{equation}
where ${\bm{T}} (z)$ is the vector of the Chebyshev polynomials $T_k$ at $z$.

Integrals (8) and (11) are to be computed at $z = 1$, therefore, we obtain from equation (E4) 
   \begin{equation}
      \int_0^1 f(\xi) \, d \xi = {\bm{P}} \cdot {\bm{C}} (f),
   \end{equation}
where
   \begin{equation}
      {\bm{P}} = {\bm{T}} (1) \, {\bm{J}},
   \end{equation}
and the vector ${\bm{T}} (1)$ is given by equation (B4b).

%\bibliography{/Users/VG/Dropbox/VG-File-Exchange/DCP/text/references/refs-vg-DCP} % Mac
%\bibliography{C:/Users/erf/Dropbox/VG-File-Exchange/DCP/text/references/refs-vg-DCP} % MRO PC 
%\bibliography{./refs-vg-DCP} % local

\end{document}